\documentclass[12pt]{article}
\newcommand{\ft}[2]{{\textstyle\frac{#1}{#2}}}
\def\Re{\mathop{\rm Re}\nolimits}

\def\rme{{\rm e}}
\def\rmi{{\rm i}}
\def\rmd{{\rm d}}
\newsavebox{\uuunit}
\sbox{\uuunit}
    {\setlength{\unitlength}{0.825em}
     \begin{picture}(0.6,0.7)
        \thinlines
        \put(0,0){\line(1,0){0.5}}
        \put(0.15,0){\line(0,1){0.7}}
        \put(0.35,0){\line(0,1){0.8}}
       \multiput(0.3,0.8)(-0.04,-0.02){12}{\rule{0.5pt}{0.5pt}}
     \end {picture}}

\csname @addtoreset\endcsname{equation}{section}

\begin{document}

\begin{titlepage}
\begin{flushright}
SU-ITP-03/32\\
NYU-TH-03/11/09\\
KUL-TF-03/33\\
hep-th/0312005
\end{flushright}
\vspace{.5cm}
\begin{center}
\baselineskip=16pt
{\bf \LARGE $D$-term strings
}\\
\vfill
{\large Gia Dvali $^1$, Renata Kallosh $^2$ and Antoine Van Proeyen
$^3$ } \\
\vfill
{\small $^1$
Center for Cosmology and Particle Physics, \\ [1mm]
Department of Physics, New York University,\\
New York, NY 10003\\ \vspace{6pt}
$^2$ Department of Physics, Stanford University,\\
Stanford, CA 94305-4060, USA.\\ \vspace{6pt}
$^3$ Instituut voor Theoretische Fysica, Katholieke Universiteit Leuven,\\
       Celestijnenlaan 200D B-3001 Leuven, Belgium.
 }
\end{center}
\vfill
\begin{center}
{\bf Abstract}
\end{center}
{\small We study the embedding of  cosmic strings, related to the
Abrikosov-Nielsen-Olesen vortex solution,  into $d=4$, $N=1$
supergravity. We find that the local  cosmic string solution which
saturates the BPS bound of supergravity with $D$-term potential for the
Higgs field and with constant Fayet--Iliopoulos term, has 1/2 of
supersymmetry unbroken. We observe an interesting relation between the
gravitino supersymmetry transformation, positive energy condition and the
deficit angle of the cosmic string. We argue that the string solutions
with magnetic flux with $F$-term potential cannot be supersymmetric,
which leads us to a conjecture that D$_1$-strings (wrapped
$D_{1+q}$-branes) of string theory in the effective $4d$ supergravity are
described by the $D$-term strings that we study in this paper. We give
various consistency checks of this conjecture, and show that it
highlights some generic properties of non-BPS string theory backgrounds,
such as brane-anti-brane systems.  Supersymmetry breaking by such systems
can be viewed as FI $D$-term breaking, which implies, under certain conditions, the presence of
gauged $R$-symmetry on such backgrounds. The
$D$-term nature of the brane-anti-brane energy can also provide
information on the superpotential for the tachyon, which Higgses the
$R$-symmetry. In this picture, the inter-brane force can be viewed as a
result of the world-volume gauge coupling renormalization by the open
string loops.

 }\vspace{2mm} \vfill \hrule width 3.cm
{\footnotesize \noindent e-mails: gd23@feynman.acf.nyu.edu,
kallosh@stanford.edu, antoine.vanproeyen@fys.kuleuven.ac.be }
\end{titlepage}
\tableofcontents{}
\section{Introduction}

Some important issues in $d=4$ $N=1$ supergravity with constant
Fayet--Iliopoulos (FI) term are currently clarified in \cite{I} on the
basis of the superconformal approach to supergravity
\cite{Barbieri:1982ac,Ferrara:1983dh,Kallosh:2000ve}. It is natural
therefore to look  for some extended objects in this theory that have a
fraction of unbroken supersymmetry. A  candidate for such a
supersymmetric solution is a BPS cosmic string in the critical
Einstein-Higgs-Abelian gauge field model~\cite{book}.  Such BPS solutions
have been identified in \cite{Comtet:1988wi} long time ago: they are
known to saturate the Bogomol'nyi bound. In the presence of a single
charged scalar field, the vacuum manifold has non-trivial homotopy
$\pi_1\, = \, Z$, and the theory possesses topologically non-trivial
string configurations that carry $U(1)$-magnetic flux.

However, so far these solutions have not been embedded into supergravity
and therefore it was not clear what kind of unbroken supersymmetry in
what kind of supergravity explains the saturation of the bound for the
cosmic strings. We will show here that cosmic strings have 1/2 of
unbroken supersymmetry when embedded into an $N=1$, $d=4$ supergravity
with constant FI terms\footnote{In \cite{Gutowski:2001pd} a class of
string solutions of supergravity with constant FI term was found, with
1/2 of unbroken supersymmetry. Their model  describes axially symmetric
solutions without the charged Higgs field $\phi$. Therefore these
solutions are not localized near the core of the string on the scale of
the inverse mass of the vector field $m\sim g \langle \phi \rangle$, as
in the usual local cosmic strings that we are going to describe in our
paper.} in a model that has a $D$-term potential for the charged Higgs
field. Such Higgs field acquires a vev away from the core of the string,
which compensates the FI term. This provides a  $D$-flatness condition,
$D=g\xi-g\phi^* \phi=0$, away from the core, for the cosmic string in the
Einstein-Higgs-Abelian gauge field model \cite{book}. This solution away
from the core was studied in (2+1)-dimensional supergravity
in~\cite{Becker:1995sp,Edelstein:1996ba}, where it was also established
that the configuration has  1/2 of unbroken supersymmetry\footnote{We are
grateful to J. Edelstein for informing us about these papers.}. However,
so far these solutions have not been embedded into 3+1 supergravity and
therefore it was not clear what kind of unbroken supersymmetry in what
kind of supergravity explains the saturation of the bound for the cosmic
strings.

We argue that in $N=1$,  $d=4$ supergravity,  strings produced by Higgs
fields that minimize  $F$-terms ($F$-term strings) cannot be BPS
saturated. Hence, the only BPS saturated strings in $N=1$, $d=4$
supergravity are $D$-term strings.

 On the other hand, it is known that string theories admit various BPS-saturated
string-like objects in the effective $4d$ theory. These are
$D_{1+q}$-branes wrapped on some $q$-cycle.  We shall refer to these
objects as effective D$_1$-strings, or D-strings for short. Then an
interesting question arises. How are the string theory D-strings seen
from the point of view of $4d$ supergravity?   If D-strings admit a
low-energy description in terms of supergravity solitons, then the only
candidates for such solitons are $D$-terms strings, since, as we shall
show, they are the only BPS saturated strings  in $4d$ theory. Thus, we
conjecture that the string theory D-strings (that is, wrapped
D$_{1+q}$-branes) are seen as $D$-terms strings in $4d$ supergravity.  We
shall give various consistency checks on this conjecture, and show that
it passes all the tests.

 Other than serve as an intriguing connection between D-branes and
supergravity solitons, our conjecture sheds new light on the properties
of some non-BPS string theory backgrounds, such as, for instance,
brane-anti-brane backgrounds that break all the supersymmetries. It has
been suggested \cite{Sen:1998ii} that a BPS D$_{1+q}$ brane can be viewed
as a tachyonic vortex formed in the annihilation of a non-BPS
D$_{3+q}-\bar{\mbox{D}}_{3+q}$ pair. According to this picture, the
annihilation can be described as condensation of a complex tachyon, which
is a state of an open string (connecting brane and anti-brane). The
tachyon compensates the energy of the original brane-anti-brane system,
and the tachyonic vacuum is the closed string vacuum with no branes and
no open strings.  The tachyon condensate Higgses  the world-volume $U(1)$
group, and there are topologically non-trivial vortex solutions that are
identified with stable BPS D$_{1+q}$-branes.
 Our conjecture puts the above picture in the following light.
 Since according to our conjecture D$_{1+q}$ branes are  $D$-term strings, it immediately follows
that the energy of the D$_{3+q}-\bar{\mbox{D}}_{3+q}$-system must be seen
from the point of view of the $4d$ supergravity as $D$-term energy, with
FI $D$-term determined by $D_{3+q}$ brane tension. Existence of a
non-zero FI term implies that  the $U(1)$-symmetry Higgsed by the tachyon
is seen on the $4d$ supergravity side as gauged $R$-symmetry. We will
raise here the issues related to the gauged $R$-symmetry in supergravity
with constant FI term.

 In short, some
parts of our correspondence can be summarized as follows:
\begin{tabbing}
 \phantom{.}\hspace{8mm}\= Ramond-Ramond-charge~of~$D$-string \
 \=$\leftrightarrow$\ \ \=\kill
 \> Wrapped~D$_{1+q}$-branes\>$\leftrightarrow$\>$D$-term strings\\
 \> Energy of  D$_{3+q} -\bar{\mbox{D}}_{3+q}$-system \> $\leftrightarrow$ \> FI
$D$-term \\
\> Open~string~tachyon \>$\leftrightarrow$\> FI-canceling Higgs\\
\> $U(1)$-symmetry~Higgsed~by~tachyon \> $\leftrightarrow$ \> gauged~$R$-symmetry\\
\> Ramond-Ramond-charge~of~$D$-string~\>$\leftrightarrow$\>
topological~axion~charge~of~$D$-term~string
\end{tabbing}

 Finally, we suggest that the proposed picture may give useful information
about the form of the tachyon superpotential. The $D$-term nature of
brane-anti-brane energy indicates that the tachyon must have a partner, a
chiral superfield of opposite $U(1)$-charge. By mixing with this
superfield in the superpotential, the tachyon gets a positive mass$^2$
term when branes are far apart. In this picture, the inter-brane
interaction potential should be interpreted as the  renormalization of
$D$-term energy (due to renormalization of the world-volume gauge
coupling) via the one-loop open string diagram.

\section{Unbroken Supersymmetry in Supergravity and BPS equations}
\label{ss:BPS}

The supergravity model is defined by one scalar field $\phi$, charged
under $U(1)$, with $K=\phi^* \phi$ and superpotential $W=0$, so that we
reproduce the supergravity version of the cosmic string in the critical
Einstein-Higgs-Abelian gauge field model~\cite{book}. This model can be
also viewed as a $D$-term inflation model \cite{combDterminfl} in which
the second charged field is vanishing everywhere at the inflationary
stage as well as at the exit stage, which leads to a vanishing
superpotential. In such case, the bosonic part of the supergravity action
is reduced to
\begin{eqnarray}
e^{-1}{\cal L}_{\rm bos}=-\ft12M_P^2 R -\hat {\partial }_\mu
\phi\, \hat {\partial }^\mu \phi^*
 -\ft{1}{4} F_{\mu \nu } F^{\mu \nu }- V^D\,,
 \label{bosonic2}
\end{eqnarray}
where the D-term  potential is defined by
\begin{equation}
V^D= \frac{1}{2} D^2 \qquad  D\, = g\xi - g \phi^*\, \phi \,.
\label{Pphi}
\end{equation}
Here $W_\mu$ is an abelian gauge field,
\begin{equation}
F_{\mu\nu}\equiv {\partial }_\mu W_\nu- {\partial }_\nu W_\mu\, ,\qquad
\hat {\partial }_\mu \phi\equiv ({\partial }_\mu -\rmi g W_\mu) \phi\,.
\end{equation}

The transformation rules of the fermions\footnote{The detailed
supersymmetry transformation for the  general class of models will be
presented in \cite{I}.} in a model with the vanishing superpotential and
trivial kinetic function for the vector multiplets are
 \begin{eqnarray}
\delta \psi _{\mu L}  & = & \left( \partial _\mu  +\ft14 \omega _\mu
{}^{ab}(e)\gamma _{ab}
+\ft 12\rmi A_\mu^B \right)\epsilon_L \,, \nonumber\\
\delta \chi_L & = & \ft12(\not\! {\partial }-\rmi g \not\! W) \phi
\epsilon _R\,,
 \nonumber\\
\delta \lambda  &=&\ft14\gamma ^{\mu \nu } F_{\mu \nu }\epsilon
+\ft12\rmi \gamma _5  D
 \epsilon  \,. \label{susyf}
\end{eqnarray}
Here the gravitino $U(1)$ connection  $A_\mu ^B$ plays an important role
in the gravitino transformations.
 \begin{eqnarray}
  A_\mu ^B&=&\frac{1}{2M_P^2}\rmi\left[ \phi\partial _\mu \phi^* -\phi^* \partial _\mu \phi\right]
  +\frac{1}{M_P^2}W_\mu  D\nonumber\\
  &=&\frac{1}{2M_P^2}\rmi\left[ \phi\hat{\partial} _\mu \phi^* -\phi^* \hat{\partial} _\mu \phi\right]
  +\frac{g}{M_P^2}W_\mu  \xi  \,.
  \label{AmuBinphi}
\end{eqnarray}
We are looking for the configuration for which  all variations of
fermions in (\ref{susyf}) are vanishing for some non-vanishing
$\epsilon$, i.e. $\delta \psi _{\mu}=0$ , $\delta \chi=0$ and  $\delta \lambda =0$. We choose the following bosonic configuration:
\begin{equation}
\phi (r,\theta )\, = \, f(r)\,{\rm e}^{\rmi n \theta}\,,
 \label{stringhiggs}
\end{equation}
where $\theta$ is an azimuthal angle, and $f(r)$ is a real function that
outside the string core approaches the vacuum value $f^2=\xi$ for which
the $D$-term vanishes. The plane of the string is parametrized by $r$ and
$\theta $, where $x=r\cos \theta$ and $y=r \sin \theta$. We consider a
metric
\begin{equation}
  \rmd s^2= -\rmd t^2 +\rmd z^2+\rmd r^2 + C^2(r) \rmd \theta ^2\,.
 \label{tentativemetric}
\end{equation}
We use the vielbein with $e^1=\rmd r$ and $e^2=C( r)\rmd\theta $, which
leads to
\begin{equation}
  \omega_r{}^{12}=0\,,\qquad
  \omega_\theta {}^{12}= -C'(r)\,.
 \label{12components}
\end{equation}
For the gauge potential we take
\begin{eqnarray}
 g W_\mu\, \rmd x^\mu = n\alpha (r) \,\rmd\theta \,
  \qquad \rightarrow \qquad
  F=\ft12F_{\mu \nu }\,\rmd x^\mu \,\rmd x^\nu ={n \alpha '(r)\over g} \rmd
  r\,\rmd\theta = {n \alpha '(r)\over  g C(r)} e^1 e^2 \,.
 \label{explicitW}
\end{eqnarray}
To solve the equations for the Killing spinor we will use the following
projector
\begin{equation}
  \gamma ^{12}\epsilon\, = \, \mp \rmi \gamma _5 \epsilon\,.
 \label{projectioneps12}
\end{equation}

The BPS equations that are the vanishing of the transformations
rules~(\ref{susyf}) for the chiral field $\chi$ and for the gaugino
$\lambda$ are in $r,\theta $ coordinates
\begin{eqnarray}
&&\Bigl( C(r){\partial}_r \, \pm \, {\rm i} \, ({\partial }_\theta-\rmi g W_{\theta} ) \Bigr) \phi \, = \, 0\,, \nonumber\\
 &&F_{12 } \, \mp   D\, = \, 0\,, \label{bpseqrtheta}
\end{eqnarray}
which leads  to
\begin{eqnarray}
    C (r)f'(r)&=&\pm n f(r)\left[ 1-\alpha (r)\right], \nonumber\\
    \frac{\alpha '(r)}{g C (r)}&=& \pm \frac{g}{n} \left[ \xi - f^2(r)\right].
 \label{explicitBPSrho}
\end{eqnarray}
The gravitino BPS equation is now
\begin{equation}
  \partial _r\epsilon =0\,,\qquad \left[ \partial _\theta -\ft12 C'(r) \gamma
  _{12}+\ft12\rmi A_\theta ^B\right] \epsilon _L(\theta)=0\,,
 \label{gravitinoBPS}
\end{equation}
where
\begin{equation} A_r^B=0\,,\qquad
  M_P^2 A_\theta ^B =nf^2(r)+\frac{n}{g}\alpha (r) D =n \xi \alpha (r)
  +n\,f^2\left[ 1- \alpha (r)\right].
 \label{explicitAB}
\end{equation}
Using~(\ref{explicitBPSrho}), the value of $A_\theta ^B$ is
\begin{equation}
  A_\theta ^B=\frac{n}{M_P^2}\left[ \xi \alpha (r)-\frac{C (r)}{2 g^2}
  \left( \frac{\alpha '}{C (r)}\right)'\right]
 \label{AthetaBinrho}
\end{equation}
A globally well-behaving spinor parameter is
\begin{equation}
  \epsilon _L(\theta) = \rme^{\mp\ft12\rmi\theta } \epsilon _{0L}\,,
 \label{epsilontheta}
\end{equation}
where $\epsilon _{0L}$ is a constant satisfying the projection
equation~(\ref{projectioneps12}). Thus to solve the gravitino equation $\delta \psi=0$ we have to request that
\begin{equation}
  1-C'(r) = \pm A_\theta ^B\,.
 \label{diffeqrho}
\end{equation}

\section{Limiting cases}

In the limiting case where $f(r)$ is a constant, we also find that
$\alpha (r)$ is constant and the BPS equations give the values
\begin{equation}
 f^2=\xi \,,\qquad D=0\,,\qquad  \alpha =1\,,\qquad A_\theta ^B= \frac{n\xi
  }{M_P^2}\,,\qquad  C(r) =r\left(1\mp \frac{n\xi }{M_P^2}\right)\,.
 \label{constlimit}
\end{equation}
This is a cosmic string solution of  the Einstein-Higgs-Abelian gauge
field model \cite{book} far away from the core of the string at large
$r$.
\begin{equation}
\phi= \sqrt{\xi }\rme^{\rmi n\theta}\, , \qquad  W={n\over g}
\,\rmd\theta \,, \qquad F=0\,. \label{largerConfig}
\end{equation}
This solution shows that asymptotically the string creates a locally-flat
conical metric with an angular deficit angle\footnote{From unbroken
supersymmetry alone we would have $C= r \left(1\mp \frac{n\xi
}{M_P^2}\right)$. It will be explained in the next section why the sign
choice $\pm$ is equal to the sign of $n$, and therefore only the choice
$-n$ for positive $n$ and $+n$ for negative $n$ are  permitted by the
positivity of   energy.} which is due to the constant FI term $\xi$:
\begin{equation}
  \rmd s^2= -\rmd t^2 +\rmd z^2+\rmd r^2 + r^2\left(1\mp \frac{n\xi }{M_P^2}\right)^2 \rmd \theta ^2\,.
 \label{metricFAR}
\end{equation}
Notice also that the  limit $r \rightarrow\infty$   the full
supersymmetry is restored because $F_{\mu\nu}=0$, $D=0$,  $\partial_r
\phi= \hat\partial_\theta \phi= 0$ and and $R_{\mu\nu}{}^{ab}=0$ which
corresponds to  the enhancement of supersymmetry away from the core of
the string.

The other limiting case describing the string at small $r$ is
\begin{equation}
  f=0\,,\qquad D=g\xi \,,\qquad \alpha '=\pm \frac{g^2 \xi}{n} C (r)\,,\qquad 1-C'(r)
  \mp n M_P^{-2}\xi \alpha(r) =0\,,
 \label{limitf0}
\end{equation}
which can be solved with two integration constants $a$ and $b$ as
\begin{eqnarray}
 C (r) & = & \frac{M_P}{\xi g }\left[a\sin\left( g\xi M_P^{-1}r\right)+b\cos\left( g\xi M_P^{-1}r\right)\right],  \nonumber\\
  \alpha (r) & = & \pm \frac{M_P^2}{n\xi }\left[ 1-a\cos\left( g\xi M_P^{-1}r\right)
  +b\sin\left( g\xi M_P^{-1}r\right)\right].
 \label{solvef0}
\end{eqnarray}
For the vortex centre, it is required that at $r=0$, we have
$\alpha(0)=C(0)=0$. This gives that $a=1$ and $b=0$, and the solution is
thus:
\begin{eqnarray}
 \alpha (r) & = & \pm \frac{M_P^2}{n\xi }\Bigl(1-\cos( {g\xi\over M_P} r)\Bigr),  \nonumber\\
C (r) & = & {M_P\over \xi g} \sin ( {g\xi\over M_P} r)
 \label{solvef0a}
\end{eqnarray}
Notice that we have also $C'(0)=1$, and the metric at small $r$ is
non-singular:
\begin{equation}
  \rmd s^2= -\rmd t^2 +\rmd z^2+\rmd r^2 + \left[{M_P\over \xi g} \sin ( {g\xi\over M_P} r)\right]^2 \rmd \theta ^2\,.
 \label{metricCORE}
\end{equation}
At small $r$ the solution is
\begin{equation}
\phi= 0\, , \qquad  W_\theta=\pm \frac{M_P^2}{g\xi }\Bigl(1-\cos(
{g\xi\over M_P} r)\Bigr) \,, \qquad F_{r\theta}=\pm M_P \sin ( {g\xi\over
M_P} r)\,.
\end{equation}
In this limit the spacetime curvature is not vanishing,  $R= 2C''/ C = -
2\left( g \xi / M_P \right) ^2$ and half of the  supersymmetry is broken
as one can see, e. g.  from the integrability condition for the existence
of the Killing spinor.

\section{Energy and BPS conditions}

The energy of the string is (where the sums over $\mu ,\nu $ run only
over $r,\theta $ only)
\begin{eqnarray}
{\cal \mu}_{\rm string}&=& \int \,\sqrt{\det g}\,\, \rmd r \rmd\theta \left[
(\hat{\partial}_\mu \phi^*)(\hat{\partial }^\mu \phi) + \frac{1}{4}
 F_{\mu \nu }\, F^{\mu \nu} +\frac{1}{2} D^2 +\frac{M_P^2}{2}R\right] \nonumber\\
&&+\left.M_P^2\left( \int \rmd\theta \sqrt{\det h}\,\,K\right|_{r=\infty
}-\left.\int \rmd\theta \sqrt{\det h}\,\,K\right|_{r=0}\right)  \,,
 \label{estring}
\end{eqnarray}
where $K$ is the Gaussian curvature at the boundaries (on which the metric is $h$). These boundaries are at $r=\infty $ and $r=0$. For the metric~(\ref{tentativemetric}):
\begin{equation}
  \sqrt{\det g}=C(r)\,,\qquad \sqrt{\det g}\,R=2C''\,,\qquad \sqrt{\det h}\, K=-C'
\label{gRK}
\end{equation}

Eq.(\ref{estring}) can be rewritten by using the Bogomol'nyi method as
follows
\begin{eqnarray}
{\cal \mu}_{\rm string}&=& \int \, \rmd r\rmd\theta\, C(r)\left\{  \,
 |(\hat{\partial}_r \phi  \, \pm \, {\rm i}C^{-1} \, \hat{\partial }_\theta ) \phi|^2 \, + \,
{1 \over 2}\left[ F_{12} \, \mp   D
\right]^2 \right\}  \,+\label{ebpsstring}\\
&+& M_P^2\int \rmd r \rmd\theta \,  \left[\partial _r\left( C'\pm
A_\theta\right) ^B\mp\partial _\theta A_r^B\right]-M_P^2\left.\int
\rmd\theta\, C'\right|_{r=\infty }+M_P^2\left.\int \rmd\theta\,
C'\right|_{r=0}, \nonumber
\end{eqnarray}
where we have used the explicit form of the
metric~(\ref{tentativemetric}). The first line vanishes by the BPS
equations~(\ref{bpseqrtheta}). The gravitino BPS
equation~(\ref{diffeqrho}) implies that the first term in the second line
in (\ref{ebpsstring}) vanishes. The energy is thus given by the
difference between the boundary terms at $r=0$ and at $r =\infty$.  Using
eqs. (\ref{metricFAR}) and  (\ref{metricCORE})  we find
\begin{equation}
  \mu _{\rm string}=2\pi M_P^2\left(\left.C'\right|_{r=0 }-\left.C'\right|_{r=\infty }\right)= \pm 2\pi n \xi \, .
\label{energy}
\end{equation}
To have the energy positive, $ \mu _{\rm string}>0$,  we have to require,
in addition to unbroken supersymmetry,  that the metric has to be taken
in the form $C= r \left(1- \frac{|n|\xi }{M_P^2}\right)$. From unbroken
supersymmetry alone we would have $C= r \left(1\mp \frac{n\xi
}{M_P^2}\right)$. From consideration of the positivity of energy we
deduce that the deficit angle has to be positive and therefore only the
choices $-n$ for positive $n$ and $+n$ for negative $n$ in $C(r)$ are
possible.

This situation with positive deficit angle  has an exact analogy with
supersymmetric charged black holes: the mass of the black hole $M$ can be
positive or negative, while the condition of unbroken supersymmetry,
$\delta \psi=0$ can be satisfied. However, the ADM mass of the black hole
has to be positive and therefore only the case of $M= |Z| >0$, where
$|Z|$ is the absolute value of the central charge, is physical. The same happens for cosmic
strings: only positive deficit angle means positive energy of the
configuration, $\mu _{\rm string}= 2\pi |n |\xi$.

The energy of the string, which may  also defined as in ~\cite{book}
\begin{equation}
 \mu _{\rm string}=  \int \, \rmd r\rmd\theta  \sqrt{\det g}T_0{}^0 = 2\pi |n|
 \xi\,,
\label{vilenkin}
\end{equation}
where  \cite{Comtet:1988wi}
\begin{eqnarray}
T_0{}^0= \left\{  \,
 |(\hat{\partial}_r \phi  \, \pm \, {\rm i}C^{-1} \, \hat{\partial }_\theta ) \phi|^2 \, + \,
{1 \over 2}\left[ F_{12} \, \mp   D \right]^2 \right\} \pm  M_P^2
\left[\partial _r   A_\theta ^B -\partial _\theta A_r^B\right],
 \label{T00}
\end{eqnarray}
Note that if we use for the definition of the energy only the energy of
the non-gravitational fields $\phi$ and $W_\mu$, the role of the
gravitino supersymmetry transformation remains obscure. The contribution
to the energy comes from the surface term $M_P^2  \left[\partial _r
A_\theta ^B -\partial _\theta A_r^B\right]$ and is due to the vector
field, which at large $r$ is defined by  $A_\theta^B=
\frac{1}{2M_P^2}\rmi\left[ \phi\hat{\partial} _\theta \phi^* -\phi^*
\hat{\partial} _\theta\phi\right]
  +\frac{g}{M_P^2}W_\theta  \xi = \frac{n\xi
  }{M_P^2}$. The first two terms in (\ref{T00}) vanish due to saturation of the BPS bounds, $(\hat{\partial}_r \phi  \, \pm \, {\rm i}C^{-1} \, \hat{\partial }_\theta ) \phi=0$ and $F_{12} \, \mp   D=0$. These two bounds are saturated since the supersymmetry variation of the chiral fermion $\chi$ and  gaugino $\lambda$ has to vanish.

The definition of the energy of the string that we are using in
(\ref{estring}), which is valid for time independent configurations, is
\begin{equation}
  E= \int_M \sqrt {\det g} \,\left(\frac{M_P^2}{2}R -L_{\rm matter}\right) + M_P^2\int_{\partial M} \sqrt {\det h}\,
  K\,.
\label{ENERGY}
\end{equation}
Now we see that the term $\left(\frac{M_P^2}{2}R -L_{\rm matter}\right)$
produced in addition to two BPS bounds in (\ref{T00}) also a term
$\left[\partial _r\left( C'\pm A_\theta\right) ^B\mp\partial _\theta
A_r^B\right]$. Due to the gravitino BPS bound,~(\ref{diffeqrho}), the
surface term $\partial _r   A_\theta ^B$ in $T_0{}^0$ is cancelled by the
Einstein term $\sqrt g \, R$. This is not surprising since the Einstein
equation of motion must be satisfied due to vanishing gravitino
transformations. The remaining term in the energy, the Gibbons-Hawking
$K$ surface term, gives the non-vanishing contribution to the energy of
the string which is directly related to the deficit angle $\Delta$, where
$M_P^2 \Delta = \mu_{\rm string}$.

\section{Are $D$-term strings the D-strings?}

 Note that in $N=1$ supersymmetric theories, one could easily come up with
non-trivial superpotentials such that the minimization of $F$-term(s)
triggers the spontaneous breaking of some gauge $U(1)$-symmetries. Such
theories would also admit topologically non-trivial string solutions,
which we shall refer to as the $F$-term strings. In the core of the
$F$-term strings some of the $F$-terms become non-zero.
  It is important to note that $F$-term strings, as opposed to $D$-term strings
{\it cannot} be BPS saturated objects, in the sense that they break all
the supersymmetries. This fact can be directly understood from the
gaugino transformation on the string background. Indeed, the only way to
compensate the magnetic contribution to gaugino variation, is through the
non-zero $D$-term part. This is impossible if the potential energy
contribution to the string mass comes  from $F$-terms.
 Let us briefly demonstrate the above statement on a simple example.
We shall take a globally supersymmetric theory, with $U(1)$ gauge group
spontaneously broken by two chiral superfields $\phi_+, \phi_-$ with
opposite $U(1)$ charges. The relevant superpotential is:
\begin{equation}
 W \, = \,  \lambda X (\,\phi_+\phi_- \, - \, \eta^2)\,,
\label{fsuper}
\end{equation}
where, $\lambda$ and $\eta^2$ are positive constants and $X$ is a
$U(1)$-neutral chiral superfield. Generalization of this example to the
supergravity case is straightforward and adds nothing to our conclusion.
The potential energy of this system is
\begin{eqnarray}
V &=& V_F + V_D \ , \nonumber \\
V_F &=& \lambda^2 |\phi_+\phi_-\, - \, \eta^2|^2 \,  + \, \lambda^2
|X|^2|\phi_-|^2  \, + \, \lambda^2 |X|^2|\phi_+|^2.
\nonumber \\
V_D &=& {g^2 \over 2} \left[|\phi_+|^2 \, - \, |\phi_-|^2 \right]^2\,,
\label{D+F}
\end{eqnarray}
which is minimized for $\phi_+ \, = \, \phi_- \, = \eta $ and $X \, = \,
0$, and with unbroken supersymmetry and spontaneously broken $U(1)$ gauge
group. Hence there are topologically stable strings in this model.
Because of the opposite charges, $\phi_+$ and $\phi_-$ will have opposite
winding of the phase around the string, that is the string solutions will
have a form $\phi_\pm = f_\pm(r){\rm e}^{\pm \rmi\theta}$. Let us study
the energy and the supersymmetric properties of these objects. First, we
shall note that in the lowest energy configuration $|X|$ must vanish, as
it only multiplies positive definite terms in the potential. Hence, we
can simply set $X\, = \,0$.

Next, we note that because the system is invariant under the exchange
$\phi_+ \rightarrow \phi_-^*$, and because the $D$-term is minimized for
$|\phi_-| = |\phi_+|$, the lowest possible energy string configuration
must have $f_+ = f_-$. Hence, we can set  $\phi_+ = \phi_-^*$. The
non-zero part of the string energy can only satisfy the Bogomol'nyi limit
provided we set $\lambda^2 = g^2/2$. With this choice of the parameters
and with the above ansatz, the energy of the scalar fields becomes
identical to the one given by (\ref{bosonic2}), where the role of $\xi$
is played by $2\eta^2$. We have shown above that such a system does have
the string solution that energetically satisfies the BPS bound.

Although it formally satisfies the energy bound, nevertheless, such a
configuration breaks all the supersymmetries. To see this, it is enough
to check the transformation properties of the gaugino, and the chiral
fermionic partner of $X$ (we shall call it $\chi$) in the string
background:
\begin{eqnarray}
\delta \chi_L& = & - \ft12  F_X \epsilon _L\,,
 \nonumber\\
\delta \lambda^\alpha  &=&\ft14\gamma ^{\mu \nu } F_{\mu \nu }\epsilon\,.
\label{chilambda}
\end{eqnarray}
Here $F_X \, = \,  \lambda (\phi_+\phi_-\, - \, \eta^2)$ is the $F$-term
of $X$. Because, both $F_{\mu \nu }$ as well as $F_X$ are non-zero in the
string core, (with $F_{\mu \nu }$ of magnetic type, $\gamma ^{\mu \nu
}F_{\mu \nu }$ is invertible) the above variations cannot vanish for any
choice of $\epsilon$, and hence $F$-term strings break all the
supersymmetries. Even if the parameters are fine tuned in such a way that
at the tree level such strings satisfy the Bogomol'nyi energetic bound,
nevertheless they are not BPS saturated objects in the sense of unbroken
supersymmetry. Since all the supersymmetries are broken, the tree level
Bogomol'nyi bound is not protected against quantum corrections and in
general will be destabilized in perturbation theory, since the physical
gauge and scalar couplings are renormalized differently. Hence, the
choice $\lambda^2 = g^2/2$ is not perturbatively stable.

  In supergravity the story is very similar.  Let us choose a minimal K{\"a}hler
$K \, = \, |\phi_+|^2  \, +\,  |\phi_-|^2 \, + \, |X|^2$.  For the
$F$-term strings, in order to achieve the saturation of the Bogomol'nyi
energy bound, even in the case of a minimal K{\"a}hler, we have to allow for
a non-minimal gauge kinetic function, for reasons that will become clear
in a moment. Again, just as in rigid limit, it can be shown easily that
making either $|X| \neq 0$, or $\phi_+ \neq \phi_-^*$
 only increases the energy.
Thus, we set  $X=0$,  and $\phi_+ = \phi_-^* = \phi/\sqrt{2}$, and the
static $z$-independent energy of the system becomes
\begin{eqnarray}
E= \hat {\partial }_\mu \phi\, \hat {\partial }^\mu \phi^*
 + \, \frac{\Re   f(\phi) }{4} F_{\mu \nu } F^{\mu \nu } +
\frac{\lambda^2}4{\rm e}^{|\phi|^2 M_P^{-2}} \left( |\phi|^2 \, - 2\eta^2
\right) ^2 \,,
 \label{bos3}
\end{eqnarray}
where $\mu, \nu = 1,2$.
 Again this expression  can be simply
reduced to the energy of a single scalar field, with a FI D-term
potential with $\xi  = 2\eta^2$, provided the gauge kinetic function is
chosen to be exactly equal to
\begin{equation}
\label{gkf} \Re f(\phi)^{-1} \, = \, {\lambda^2 \over 2g^2} {\rm
e}^{|\phi|^2 M_P^{-2}}\,.
\end{equation}
This condition is analogous to the choice $\lambda^2 = 2g^2$ in the rigid
limit. Such a system admits the string solutions that satisfy Bogomol'nyi
equations. In the $\xi \ll M_P^2$ limit, the solution in leading order
coincides with our  $D$-term cosmic string solution up to corrections of
order $\xi M_P^{-2}$. Again, although the solution satisfies an energetic
BPS bound, it is {\it not} a BPS state in the sense of an unbroken
supersymmetry, since it breaks all the supersymmetries.

 To summarize, by a special choice of the K{\"a}hler and gauge kinetic function, the  $F$-term
string energy can formally made equal to some $D$-string energy, and
thus, saturate the Bogomol'nyi  energy limit. However, the supersymmetric
properties of the two solutions are very different, due to the fact that
the participating fields belong to the different multiplets. $F$-term
strings break all the supersymmetries, whereas the $D$-term strings
preserve half of it. This is also reflected in the fact that the choice
(\ref{gkf}) is not stable under radiative corrections, and hence the
bound is not expected to be maintained to all orders in perturbation
theory, whereas for the $D$-term strings, the equality of the gauge
coupling to the  Higgs self-coupling is imposed by supersymmetry.

 It is obvious that the addition of an FI term cannot change this
situation, as long as $\lambda \neq 0$. First, because the exact
cancellation between the $D$-term and $F_{\mu\nu}$ in the gaugino
variation will be impossible. Secondly, for $\lambda \neq 0$, $F_X$
cannot be made identically zero throughout the solution.

Another argument showing that $F$-term strings cannot be BPS-saturated in
supergravity theories, is based on the gravitino transformation
properties in (\ref{susyf}). The crucial point, that enabled the
existence of Killing spinors on the space with a conical deficit angle,
is the conspiracy between gauge and spin connections. This was possible
only because the gravitino change under the broken $U(1)$ was set by FI
term, which in the same time sets the string tension, and thus, the
deficit angle. Hence, transporting the spinor around the string, the
phases acquired due to the $U(1)$ charge and due to the deficit angle
cancel, and the existence of the Killing spinors is
possible~\cite{Becker:1995sp}. On the other hand if the contribution into
the string tension were coming from the $F$-terms, this cancellation
would be impossible, since the $U(1)$-charge of gravitino only comes from
FI term, and is independent of the value of $F$-terms.

 Having argued that $D$-term strings are the only BPS-saturated strings in
$N=1$, $d=4$ theory, we wish to point out a possible intriguing
connection, between the $D$-term strings, and D$_1$-strings in type $II$B
string theories (see \cite{Joe1} for a review). Namely, we wish to
conjecture that the D$_1$-strings as seen from the point of view of the
low-energy supergravity are in fact $D$-term strings. This connection can
shed some new light on D-string formation in D-brane annihilation.

To explain the connection, let us recall first some of the known facts of
unstable brane-anti-brane systems~\cite{Sen:1998ii}. Consider for
simplicity a D$_p \, - \, \bar{\rm D}_p$ pair in type $II$ string theory.
We shall assume that $6$ dimensions are compactified on a torus, and the
D-branes can move relative to each other in some of them. When branes are
close to each other, the open string connecting  them has a tachyon in
the spectrum, which signals instability.  It was conjectured
\cite{Sen:1998ii} that the annihilation can be described in terms of the
tachyon condensation. The topology of the tachyonic vacuum is non-trivial
and can be described in terms of K-theory~\cite{Witten:1998cd}.

  For a single brane-anti-brane pair, the low-energy symmetry
group consists of two $U(1)$ symmetries, and the tachyon is charged under
a linear combination of these. After tachyon condensation, the diagonal
$U(1)$ subgroup gets Higgsed. The tachyon is complex, but the gauge
invariance implies that the potential energy can only depend on its
absolute value. Hence the tachyonic vacuum has a non-trivial $\pi_1$
homotopy, and there must exist stable tachyonic string solutions. These
strings must lie in the plane of the parent branes, and have
co-dimensions two less than the dimension of these branes. They are
identified with the stable BPS D$_{p-2}$ branes. In our case, since all
the dimensions are compact these objects will effectively look as D$_1$
strings from the four-dimensional point of view. The question is what
these strings correspond to as seen from the four-dimensional
supergravity. It is tempting to assume that they should correspond to
some stable BPS strings in that theory. But if so, we have argued that
the only possible BPS strings in supergravity are the $D$-term strings,
and hence they are the only candidates for D-strings.

 This fits the picture. Indeed, the energy density
of the  D$_{p} - \bar{{\rm D}}_{p}$ system breaks all the
supersymmetries. According to the tachyonic description, supersymmetry is
restored by the tachyon condensation, which cancels the  energy of
 the original system.
The tachyonic vacuum must possess stable BPS D$_{p-2}$ branes, which must
correspond to BPS strings in effective four-dimensional supersymmetric
theory. But we have shown above that the only candidate for such strings
are the $D$-term strings. Hence the initial energy of the
brane-anti-brane system compensated by the tachyon condensate must exist
in form of the $D$-term rather than the $F$-term, or else tachyonic
strings would be non-BPS states.

The Ramond-Ramond  charges of the D$_{p-2}$ branes that  in the D-brane
description come from the coupling of the $(p-1)$-form  RR field
($C_{(p-1)}$)  to the $U(1)$  magnetic flux on the  world volume of the
unstable brane-anti-brane pair
\begin{equation}
\label{rr} \int_{p+1} \, F_{(2)} \wedge C_{(p-1)}
\end{equation}
at the level of an effective $4d$ supergravity  $D$-term strings, should
be translated as charges with respect to the "axionic" multiplet, coupled
to the magnetic flux. Coupling to such a multiplet  should not undermine
the BPS properties of the  $D$-term strings. For instance the repulsion
between the two parallel strings mediated by the pseudoscalar axion
should be  exactly compensated by the attraction of its scalar partner.

\section{A consistency check: matching tensions}\label{ss:matchtensions}

 We wish to give a brief consistency check of our conjecture of $D$-term-D-string correspondence,
and show that the $D$-term string tension has the correct scaling
properties for the D-brane. Consider a D$_1$ string formed in the
annihilation of a D$_3-\bar{\mbox{D}}_3$ brane pair.  According to the
description in terms of the open string tachyon condensation, the D$_1$
brane is a tachyonic vortex, which carries the magnetic flux of the
Higgsed  diagonal $U(1)$ subgroup of the original $U(1)\times U(1)$
symmetry group.  The tachyonic vacuum, with no branes,  is supersymmetric
and the D$_1$ brane breaks half of it.

 According to our conjecture, this picture translates as follows. The D$_3-\bar{\mbox{D}}_3$ system
is seen as the background with a non-zero FI  $D$-term for the diagonal
$U(1)$. Hence, supersymmetry is broken by the presence of the $D$-term.
Supersymmetry is restored by the condensation of the Higgs field that
cancels the FI term. D$_1$-strings are just the BPS $D$-term strings
discussed above.  We shall now match the scaling properties of the
corresponding parameters involved in the problem.  For simplicity, we
shall keep the six additional dimensions uncompactified. In this limit
the effective $4d$ gravitational constant vanishes and $4d$ supergravity
approaches the rigid limit.   Matching of the D$_3-\bar{\mbox{D}}_3$
tension to the $D$-term energy implies
\begin{equation}
\label{tensions}
2T_{3} \, = \, {2 \over g_s(2\pi)^3 \alpha^{'2}} \, = \, {g^2 \over 2} \xi^2,
\end{equation}
where $g_s$ is the string coupling.  Matching of the D$_1$-brane and D-string tensions
gives the following relation
\begin{equation}
\label{stringtensions}
T_{1} \, = \,{ 1\over g_s(2\pi) \alpha' } \, = \mu_{\rm string} \, = \, 2\pi
\xi\,.
\end{equation}
The two are consistent, provided $g^2 \, = \, 8\pi g_s$. This is indeed a
correct scaling relation between string and gauge couplings. The
additional factor of two is the result of the relative charge of our
Higgs field (which we have normalized to one) with respect to the open
string tachyon, to which in general it is related by a non-trivial field
redefinition. The above matching can be generalized for BPS D-branes
formed in annihilation of different D$_p-\bar{\mbox{D}}_p$ systems. For
instance, for the  D$_{1+q}$-branes formed in the annihilation of
D$_{3+q}-\bar{\mbox{D}}_{3+q}$ branes, with $q$ world-volume dimensions
wrapped on a $q$-torus of radius $R$, we have the following matching
conditions
\begin{equation}
\label{tensions1} {2 R^q \over g_s(2\pi)^3 \alpha^\prime{}^{q+4\over 2}}
\, = \, {g^2 \over 2} \xi^2\,,
\end{equation}
and
\begin{equation}
\label{stringtensions1}
  {R^q\over g_s(2\pi) \alpha^\prime{}^{q +2 \over 2}} \, = \, \mu_{\rm string} \, = \, 2\pi
 \xi\,,
\end{equation}
which is satisfied for $g^2 = 8\pi g_s {\alpha^\prime{}^{q \over 2} \over
R^q}$.

 The fact that the tensions match so precisely is expected, but it is also surprising.
It is expected from the general scaling relation of string and gauge
couplings that tensions of the gauge theory solitons  should scale as the
ones of D-branes. On the other hand the soliton in question is derived
from the low-energy gauge theory, and one would naively expect a
correction due to the infinite tower of states integrated out. Once
again,  supersymmetry and BPS properties seem to give us much better
control of the situation.

\section{RR charges}

D$_1$-branes are charged with respect to the two-form RR field $C_{(2)}$.
If our conjecture of $D$-term-D$_1$-string correspondence is correct,
this long-range RR field must have a counterpart in the $D$-term string
description. As noted above, an obvious candidate for such a long-range
field is the axion ($a$). Indeed, the axion is dual to a two-form field,
and it is not surprising that this two-form is exactly the RR two-form
field of the D$_1$-brane. Thus, the electric RR-charge of the D$_1$-brane
translates as the topological (winding) axionic charge of the
$D$-term-string.  We wish to analyse more closely how this relation comes
about.

Let us again think about the formation of the D$_{1 + q}$-brane in D$_{3
+ q}-\bar{\mbox{D}}_{3+q}$ annihilation. As before, we assume that $q$
dimensions are wrapped on a compact cycle, so that the daughter D$_{1 +
q}$-brane effectively looks as D$_1$ brane from the point of view of $4d$
theory. The low energy symmetry group is $U(1)\times U(1)$, one linear
superposition of which is Higgsed by the tachyon VEV.
 D$_1$ strings are vortices formed by the winding of the tachyon phase, which support
the gauge flux of the Higgsed $U(1)$.  Notice that this Higgsed $U(1)$
gauge field is precisely the combination of the original $U(1)$-s that
carries a non-zero RR charge (the other combination is neutral).  The
corresponding gauge field strength ($F_{(2)}$) has a coupling to the
$C_{(2)}$ form via the WZ terms on the world-volume of the unstable
parent brane pair
\begin{equation}
\label{fccoupling} \int_{3 + 1 + q} {1 \over M_P} F_{(2)}\wedge C_{(2 +
q)}\,,
\end{equation}
where, since we are interested in the effective $4d$ supergravity
description,  we have only kept the $4d$ zero mode component of the RR
field. The above coupling ensures that the tachyonic vortex has a correct
$D_{1 + q}$ RR charge. The unit magnetic flux flowing in the $z$
direction, generates a long-range RR field, with the following asymptotic
energy density in $r$-direction:
\begin{equation}
\label{cenergy} (\rmd C_{(2)})^2|_{r \rightarrow \infty}  \, = \, {T_1^2
\over M_P^2} {1\over 4\pi^2 r^2}\,,
\end{equation}
where $T_1$ is the effective one-brane tension given by
(\ref{stringtensions1}), and $C_{(2)}$ is an effective two-form field
obtained via dimensional reduction of $C_{(2 + q)}$ in which $q$ indices
take values in compact dimensions.

This gives a log-divergent integrated energy per unit D-string length. To
make connection with the $D$-term-string language, we can go into the
dual description of the $C_{(2)}$-form in terms of an axion
\begin{equation}
\label{cduala} \rmd C_{(2)} \rightarrow *\,\rmd a\,,
\end{equation}
where star denotes a $4d$ Hodge-dual. Under this duality transformation
we have to replace
\begin{equation}
\label{c-a} (\rmd C_{(2)})^2  \rightarrow M_P^2 (\rmd a \, - \,g Q_a
W)^2\,,
\end{equation}
where $Q_a \, = \, {\xi \over M_P^2} $ is the axion charge under $U(1)$.
This charge vanishes as the compactification volume goes to infinity, and
$4d$ supergravity approaches the rigid limit. The above value of the
axionic charge for the $D$-term string, reproduces the correct RR charge
of the D-string, and also has a correct scaling for the anomaly
cancellation (see below).

 Let us show that the axionic long-range energy of the $D$-term string exactly matches
the RR long-range energy (\ref{cenergy}) of the D$_1$-brane. There are
two phases in the problem, one is the phase $\Theta$ of the
$D$-term-compensating field $\phi$, and the other one is the axion $a$.
Both are defined modulo $2\pi$. The gradient energies of these two fields
far away from the string core are given by the following terms
\begin{equation}
\label{gradenergy}\int r\rmd r \rmd\theta \, \frac{1}{r^2}\left[ \xi
(\partial_{\theta}\Theta \, - \, gW_{\theta})^2 \, + M_P^2
(\partial_{\theta}a \, - \, gQ_a W_{\theta})^2\right]  \,,
\end{equation}
where the configuration is given by~(\ref{largerConfig}).

Because the VEVs  must be single valued,  both phases must change around
the string by an (integer)$\times 2\pi$. That is, we must have
\begin{equation}
\label{windingsatheta} {1 \over 2\pi}  \oint \rmd\theta
\,\partial_{\theta } \Theta \, = \, n\,, \qquad   {1 \over 2\pi}  \oint
\rmd\theta\,
\partial_{\theta }  a \, = \, n_a\,.
\end{equation}
Hence, the $D$-term string is characterized by two integer winding
numbers $(n,\,n_a)$.  Around the minimal $D$-term string we have  $n =
1$. For $n=1$, the  energetically most favorable value of $n_a$ is
determined by the charge  of axion.  For $Q_a \neq 1$, there is no way to
compensate the gradient energy by any integer $n_a$, and hence there is a
long-range field around the string. For large $r$ the energy density is
\begin{equation}
\label{axionenergy} M_P^2{(n_a \, - n\, Q_a)^2\over r^2}\,.
\end{equation}
 In the weak $4d$ gravity limit $Q_a \ll 1$, the lowest energy configuration
with a unit flux, is characterized by the winding $n_a=0$, that is, the
axion prefers not to wind at all. Hence, the long-range energy behaves as
\begin{equation}
\label{logenergy}  {\xi^2 \over M_P^2 r^2}\,,
\end{equation}
which exactly matches the RR long-range energy in (\ref{cenergy}).

 The fact that  the $D$-term string has a long-range axionic field, is another
consistency check of our conjecture that it  is a $D_1$-string.
 The way in which the long-range field appears for the $D$-term string  is rather profound,
and naively one would not expect such a situation. We wish to go over
this effect once again and look at it from a slightly different angle.
The system exhibits two  $U(1)$ symmetries, which are the shifts of
$\Theta $ and of $a$, and thus these fields are two Goldstone particles.
One combination is the gauged $U(1)$ symmetry, under which
\begin{equation}
\Theta \rightarrow \Theta + g\alpha(x)\,,\qquad a \rightarrow a \, + \,g
Q_a \alpha(x)\qquad \mbox{i.e.}\qquad   \delta_\alpha  \Theta = g\,,
\qquad \delta_\alpha a= g Q_a=g {\xi \over M_P^2}\,.
 \label{deltaThetaa}
\end{equation}
The combination
\begin{equation} \label{goldstone}
G=\xi \Theta \,\delta _\alpha \Theta + M_P^2 a \,\delta_\alpha a
 = g\xi \left(  \Theta \, +\, a\right) \,,
\end{equation}
is eaten up by the $U(1)$  gauge field. Hence, this combination is
removed by the Higgs effect.  The remaining combination in the action is
\begin{equation}
\label{realaxion} g_{\rm axion} \, \equiv   \frac1g \left( \Theta
\,\delta _\alpha a - a \,\delta_\alpha \Theta \right) = \, {\xi \over
M_P^2} \Theta \, - \, a
\end{equation}
stays massless (apart of a possible anomaly, see below). Because both
$U(1)$-symmetries are non-linearly realized, the topology of the vacuum
manifold is $S_1\times S_1$ and one expects the two basic types of cosmic
strings:  1) gauge strings with magnetic flux and  no long-range field;
and 2) global strings with the long-range $g_{\rm axion}$-field, but no
magnetic flux. One may think that the gauge strings carry no long-range
field, because the winding of $G$ exactly compensates the vector
potential, and accordingly,  the global strings carry no flux because $G$
does not wind around them. However, this is impossible. The reason is
that it is $\Theta$ and $a$, and not $G$ and $g_{\rm axion}$ that must be
single valued around the string.  And because, $a$ carries a fractional
charge, the former requirement does not imply the latter. Hence, the
gauge strings with flux also carry a long-range field, and vice versa.
This property gives another consistency check of the conjecture that
$D$-term strings are indeed D-strings, that carry long-range RR-flux.

 Now let us briefly discuss what happens with the $D$-term strings in case
that the anomaly is canceled by the Green-Schwarz \cite{Green:1984sg}
mechanism. In such a case $a_{\rm axion}$ has a coupling to {\it all} the
gauge field strengths that have non-zero mixed anomalies
\begin{equation}
\label{axionff} g_{\rm axion}\, F\tilde{F} \, = \, \left({\xi \over
M_P^2} \Theta \, - \, a\right) F\tilde{F}\,.
\end{equation}
Note that the first term is generated by the chiral anomaly which is
canceled by the second one. The above coupling generates a  non-trivial
potential for $g_{\rm axion}$ through the instantons, which breaks the
continuous shift symmetry down to a subgroup (defined by an anomaly-free
discrete subgroup of the global $U(1)$ symmetry), and strings become
boundaries of the domain walls. Again the similar instability is known to
take place for $D$-strings in type $II$ string
theory~\cite{Copeland:2003bj}, and this is another consistency check of
our conjecture.

\section{$R$-symmetry}

Our conjecture of $D$-term-D-string correspondence leads us to the
conclusion that the $U(1)$ symmetry Higgsed by the tachyon VEV is in fact
a gauged $R$-symmetry. We shall now try to analyse this connection in
more detail.  Again, we shall take tachyon condensation in
$D_{3+q}-\bar{D}_{3+q}$ system as an example, assume that the six extra
dimension are compactified on a torus, and  keep the compactification
volume as a free parameter.

 We start discussing the infinite-volume case first.  As noted above, in this limit the $4d$ supergravity
approaches the rigid limit ($M_P \rightarrow \infty$). The
D$_{3+q}-\bar{\mbox{D}}_{3+q}$ system breaks all supersymmetries, and,
according to our picture, in the world-volume theory this breaking is
seen as the spontaneous breaking by the non-zero FI term $\xi$ given by
the equations (\ref{tensions}) and (\ref{stringtensions}).  In this
limit, obviously, the $4d$ gravitino is decoupled from the $U(1)$-gauge
field. Let us now make the volume finite.  The general expression for the
covariant derivative on the gravitino is in our
conventions~\cite{Kallosh:2000ve}
 \begin{equation}
 {\cal D}_{[\mu }\psi _{\nu ]}=\left(
\partial _{[\mu} +\ft14 \omega _{[\mu} {}^{ab}(e)\gamma _{ab} +\ft
12\rmi A^B_{[\mu} \gamma _5\right)\psi _{\nu ]}\,. \label{covdergr}
\end{equation}
where the $U(1)$-connection $A_\mu ^B$ is given by
\begin{equation}
  A_\mu ^B=\frac{1}{2}\rmi\left[ (\partial _i{\cal K})\hat \partial _\mu z^i-(\partial
  ^i{\cal K})\hat \partial _\mu z_i\right]
  +\frac{g\xi}{M_P^2}W_\mu  \,,
  \label{AmuB}
\end{equation}
where
\begin{eqnarray}
\hat \partial _\mu z_i=  \partial _\mu z_i- W_\mu\, \eta_i(z)\,.
\end{eqnarray}
Here, ${\cal K}$ is the K{\"a}hler function, and sum runs over all the chiral
superfields $z_i$ and $\eta_i(z) $ are the holomorphic functions that set
the  $U(1)$  transformations of  all chiral superfields in the
superconformal action,
\begin{equation}
  \delta  z_i= \eta_{i}(z)\alpha(x)\, .
\label{eta}
\end{equation}
In case that the K{\"a}hler potential is $U(1)$-invariant, the $U(1)$ gauge
transformation of the gravitino gauge-connection $A_\mu ^B$ takes a
universal form:
\begin{equation}
 \delta  A_\mu ^B= \frac{g\xi}{M_P^2} \delta  W_\mu = \frac{g\xi}{M_P^2} \partial_\mu \alpha(x) \, .
  \label{deltaconnection}
\end{equation}
This tells us that always when the constant FI term $\xi$ is present, the
gravitino transforms non-trivially under $U(1)$,  and hence $U(1)$ is an
$R$-symmetry.

In the example of the section~\ref{ss:BPS}, we took a minimal K{\"a}hler
${\cal K} = zz^*$, with a single superfield $z={\phi \over M_P}$, in
which case (\ref{AmuB}) reproduces (\ref{AmuBinphi}).  Now we have to
include an additional chiral superfield (call it $Y$), such that  the
imaginary part of its scalar component is the axion $Y = \sigma + \rmi
a$. The real part of $Y$ comes from some combination of fields in NS-NS
sector, we shall discuss this issue briefly below. Because of the shift
symmetry, the K{\"a}hler potential must depend on a combination $Y + Y^*$.
 We see that,  since the axion shifts under $U(1)$:
$\eta_Y \, = \, \rmi g Q_a/2$ there is an additional volume-dependent
contribution to the $D$-term, according to
\begin{eqnarray}
 D(z,z^*) = -\rmi\,M_P^2 \eta^i\partial _i{\cal K}(z,z^*)+ g\xi \,. \label{P}
\end{eqnarray}
To make things more explicit, we can take for a tree-level K{\"a}hler ${\cal
K}\, = \, {\rm log}(Y + Y^*)$ and set  $\phi=0$. Correspondingly, the
$U(1)$ connection takes the form
\begin{equation}
  A_\mu ^B=\frac{1}{2} {\partial _\mu a -g Q_a W_\mu  \over \sigma }
  +  \frac{g\xi}{M_P^2}W_\mu \,.
  \label{newAmuB}
\end{equation}
Under $U(1)$ gauge transformations the first term is invariant and the
second one transforms as shown in (\ref{deltaconnection}).

 In effective $N=1$, $d=4$ supersymmetric theory, the axionic field that
is coupled to the $D$-term string must have a scalar partner. We shall
now briefly discuss its origin on the type $II$B string theory side. For
this we can look to the pairing of the states directly in 10-dimensional
type $II$B sugra. The closed string states are classified by
representations of the spin $SO(8)$ group as follows. From the
NS$_+$-NS$_+$ sector: a scalar, a 2-form in a $28$-plet and a graviton in
a $35$-plet. From the R$_+$-R$_+$ sector: a scalar, an RR 2-form
$28$-plet, and an RR 4-form $35$-plet.

 This shows that the RR 4-form $C_{(4)}$, which has dimension $35$, is paired up
with the graviton from the NS-NS sector. Hence, the axion coupled to the
D$_{1+q}$ brane for $q=2$ should be paired up with the combination of the
volume modulus and the dilaton, since this modulus comes from the
graviton multiplet in 10-dimensions. This also shows that for $q=0$, the
axion is paired up with the NS-NS $B_{(2)}$ 2-form. This is also
consistent with the fact that under the $D$-term $U(1)$ only $C_{2}$
shifts but not $B_{(2)}$.

It can be explicitly checked \cite{Dall'Agata:2001zh} that for the
dimensional reduction of type $II$B theory on CY, when one gets $N=2$
theory in $d=4$, indeed the scalar dual to $C_{\mu\nu}$ partners with the
scalar dual to $B_{\mu\nu}$.
Together with the dilaton/axion they form a universal hypermultiplet. A
second hypermultiplet is formed by the volume modulus and its axion
partner dual to $C_{\mu\nu mn}$  and scalars dual to $C_{mn}$ and
$B_{mn}$, for the case of $h_{(1,1)}=1$ when $C_{mn}$ and $B_{mn}$ are
proportional to the K{\"a}hler form ($m,n$ denote components in extra
dimensions).

When compactification is on an orientifold, the universal hyper becomes a
dilaton-axion $N=1$ superfield since $C_{\mu\nu}$ and  $B_{\mu\nu}$ are
projected out. In the second one, only the volume and its axion partner
are left since $C_{mn}$ and $B_{mn}$ are projected
\cite{Giddings:2001yu}.

Finally, let us briefly comment on  the implications of the fact that the
axion shifts under the $U(1)$ $R$-symmetry. In the effective
four-dimensional supergravity when the constant FI term $\xi$ is present,
$R$-symmetry also implies that the superpotential has an $R$-charge which
is fixed by the gravitino charge. This leads to constraints on the
superpotential. If however, there is only a field dependent $D$-term, as
in known string theory models,  such constraints should be reconsidered.
These issues will be discussed in detail in \cite{I}.

\section[Tachyon superpotential and inter-brane potential from the $D$-term]
{Tachyon superpotential and inter-brane potential \\ from the $D$-term}

 In this section, we wish to very briefly discuss some intriguing implication
of our conjecture for understanding the tachyon superpotential and
brane-anti-brane potential. Our discussions here will be mostly
qualitative, and more details will be given elsewhere.

 In our picture, the supersymmetry breaking by the non BPS
brane-anti-brane system  corresponds to the spontaneous supersymmetry
breaking via FI $D$-term. We wish to ask now the following question. How
our picture can account for the inter-brane potential when brane and
anti-brane are far apart?  Again we shall work in the decompactified
limit, in which $4d$ gravity is rigid.  The energy density of the system
then is governed solely by the $D$-term energy (\ref{Pphi}). We wish to
understand how this energy (which naively looks constant for $\phi =0$)
accounts for the inter-brane force.  To answer  this question, we first
have to understand the low energy spectrum of the theory when branes are
far apart. First, when branes are far apart, there  is a (nearly
massless) field corresponding to their relative motion. This mode is a
combination of the lowest lying scalar modes of the open strings that are
attached to a brane or anti-brane only. When the branes are far apart,
the lowest lying scalar excitations of these strings are nearly massless,
since they correspond to Goldstone bosons of broken translations. We are
interested in the combination that corresponds to the relative radial
motion of branes.  In the $4d$ language this mode is a member of a chiral
superfield, which we shall call $X$.  The expectation value of $X$ than
measures the inter-brane separation (call it $r$) according to the
following relation
\begin{equation}
\label{X} X \, = \, M_s^2 r\,,
\end{equation}
where $M_s$ is the string scale.

 Next we, of course, have two gauged $U(1)$-symmetries. We shall only be interested in the
combination that according to our conjecture provides a non-vanishing
$D$-term.

 Let us now discuss the heavy states. Among all the heavy states, we shall
only be interested in the ones whose masses depend on inter-brane
separation, that is on $X$, in our language. The tachyon is certainly
among such states. This is because the tachyon is an open string state
that connects the brane and the anti-brane.  The mass of this stretched
open string is $M_s^2 r$. In our language, this means that the tachyon as
well as other open string states get mass from the coupling to our chiral
superfield $X$. Since according to our picture there are no non-zero $F$
terms, the non-zero mass cannot come  from the interactions in K{\"a}hler,
and should come from the superpotential.
\begin{equation}
\label{WXphib} W \, = \, X \phi\bar\phi\,.
\end{equation}
The $U(1)$-invariance and holomorphy  demand that the tachyon have a
partner, the chiral superfield with an opposite $U(1)$ charge. This
partner we shall call $\bar{\phi}$. This partner never gets a negative
mass$^2$ and so has a vanishing VEV.  Hence it does not affect our
conclusions so far.  However, the existence of the partner also provides
a non-trivial consistency check.  Indeed, if $\phi$ were the only
$U(1)$-charged superfield in low energy spectrum, the $U(1)$ symmetry
would have a chiral anomaly, which would have to be cancelled by GS
mechanism. But such a cancellation is impossible in the rigid limit,
since in this limit there is no $4d$ axion!
 Hence, the anomaly has to absent in the rigid limit, which is precisely the case because of the existence
of the partner.

 Now let us turn to the inter-brane potential. The energy of the system is given by the
$D$-term energy. This is constant at the tree-level, but not at one-loop
level. At one-loop level the gauge coupling $g^2$ gets renormalized,
because of the loops of the heavy $U(1)$-charged states. For instance,
there are one-loop contributions  from the $\phi$ and $\bar\phi$ loops.
More precisely there is a renormalization of $g^2$ due to one-loop open
string diagram, which are stretched between the brane and anti-brane.
Since the mass of these strings depend on $X$, so does the renormalized
$D$-term energy
\begin{equation}
\label{DX} V_D \, = \, {1\over 2}D^2  \, = \, {g^2_0 \over 2} ( 1 \, + \,
g^2_0f(X)) \xi^2\,,
\end{equation}
where $g_0^2$ is the tree-level gauge coupling, and $f(X)$ is the
renormalization function. Now remembering that $X$ sets the inter-brane
distance, and using~(\ref{stringtensions}) and the relation between
string and the gauge couplings,  we see that the $D$-term generates the
$D$-brane potential of the form
\begin{equation}
\label{VDbrane} V(r) \, = \, 2T_3 \left( 1 \, + \, g_sf(r)\right)\,,
\end{equation}
which has a correct scaling property for the D$-\bar{\mbox{D}}$
potential, computed directly from the string theory. In particular it
agrees with the fact that the brane-interaction potential can be
understood as either the tree-level closed string exchange or the
one-loop open string amplitude. This connection will be studied in more
detail elsewhere.

\section{Discussion}

In conclusion, we have solved the conditions for unbroken supersymmetry
in $N=1$, $d=4$ supergravity with the $D$-term potential and constant FI
term. The corresponding solution is a cosmic string described in detail
in \cite{book}. We have presented the form of the energy including the
gravitational energy, in which the role of all supersymmetry
transformations of the fermions, including the gravitino,  in saturating
the BPS bound is clearly explained.

We have also argued that $D$-term strings are the only BPS saturated
strings existing in $4d$ supergravity, and have conjectured that
D-strings in type $II$ string theory can be viewed as  $D$-term strings
from the supergravity perspective. We have provided various consistency
checks of of this conjecture, and have shown that it sheds new light on
some non-BPS string theory backgrounds, such as brane-anti-brane systems.
The tachyonic instability of such systems can now be viewed as an
instability produced by the FI $D$-term. Our correspondence also implies
under certain conditions the existence of gauged $R$-symmetry on such
backgrounds, which can provide a powerful constraint on the possible
forms of non-perturbatively generated superpotentials. These conditions
as well as a clear distinction between the situations with  constant FI
terms  and moduli-dependent $D$-terms in supergravity will be presented
in \cite{I}.

The study of the supersymmetric properties of the cosmic string solutions
in the effective four-dimensional supergravity theory, performed in this
paper, may be useful particularly, in view of the increasing interest to
the cosmic string solutions in string theory, see for example
\cite{Sarangi:2002yt,Dvali:2003zj,Copeland:2003bj}.

\medskip
\section*{Acknowledgments.}

\noindent It is a pleasure to thank P. Bin{\'e}truy, G. Gabadadze, S. Kachru, A. Linde, J. Maldacena, R. Myers, J.
Polchinski and A. Vilenkin for useful discussions. The work of R.K was
supported by NSF grant PHY-0244728. G. D. and A.V.P. thank the Stanford
Institute of Theoretical Physics   for the hospitality. The research of
G.D. is supported in part by a David and Lucile Packard Foundation
Fellowship for Science and Engineering, and by the NSF grant PHY-0070787.
The work of A.V.P. was partially supported by the European Community's
Human Potential Programme under contract HPRN-CT-2000-00131 Quantum
Spacetime and  in part by the Federal Office for Scientific, Technical
and Cultural Affairs, Belgium, through the Inter-university Attraction
Pole P5/27.

$~~~$ \noindent  Before we posted the revised version of this work, the
paper by E. Halyo  appeared \cite{Halyo:2003uu}, which has overlap with
section~\ref{ss:matchtensions} of our work.

\providecommand{\href}[2]{#2}\begingroup\raggedright\endgroup

\end{document}